# Twin Worlds, Divergent Fates: How Obliquity has differently shaped Pluto's and Triton's landscapes and climates

Tanguy Bertrand[1], François Forget[2], Emmanuel Lellouch[1]

[1] Laboratoire d'Études Spatiales et d'Instrumentation en Astrophysique (LESIA), Observatoire de Paris, Université PSL, CNRS, Sorbonne Université, Univ. Paris Diderot, Sorbonne Paris Cité, 5 place Jules Janssen, 92195 Meudon, France

[2] Laboratoire de Météorologie Dynamique, IPSL, Sorbonne Universités, UPMC Université Paris 06, CNRS, Paris, France

Corresponding author: Tanguy Bertrand. Email: tanguy.bertrand@obspm.fr

## Abstract

**Triton and Pluto are believed to share a common origin, both forming initially in the Kuiper Belt but Triton being later captured by Neptune (1,2). Both objects display similar sizes, densities, and atmospheric and surface ice composition, with the presence of volatile ices $N_2$, $CH_4$ and CO (3-5). Yet their appearance, including their surface albedo and ice distribution strongly differ. What can explain these different appearances? A first disparity is that Triton is experiencing significant tidal heating due to its orbit around Neptune, with subsequent resurfacing and a relatively flat surface, while Pluto is not tidally activated and displays a pronounced topography (6). Here we present long-term volatile transport simulations of Pluto and Triton, using the same initial conditions and volatile inventory, but with the known orbit and rotation of each object. The model reproduces, to first order, the observed volatile ice surface distribution on Pluto and Triton. Our results unambiguously demonstrate that obliquity is the main driver of the differences in surface appearance and in climate properties on Pluto and Triton, and give further support to the hypothesis that both objects had a common origin followed by a different dynamical history.**

# Significance Statement

Pluto and Triton are cold objects with a tenuous and condensable $N_2$ atmosphere, with traces of $CH_4$ and CO. They share the same fundamental processes of surface-atmosphere interactions and climate physics. Yet their landscapes are markedly different. We developed a numerical model able to simulate volatile transport on Pluto and Triton on the basis of straightforward physical equations. With this single model, we reproduce the observed volatile ice surface distribution on Pluto and Triton by using the same initial conditions, and we demonstrate that obliquity is the key driver of their different appearances. This tentatively supports the hypothesis that both objects had a common origin, and later a different dynamical history resulting in differences in obliquity, yielding distinct landscapes.

## Introduction

There are two main reasons supporting the fact that Triton formed in the Kuiper Belt, as Pluto did. First, the retrograde and highly inclined orbit of Triton around Neptune suggests that it was captured by the Ice Giant, probably early in solar system history (1,2). Second, Triton and Pluto have comparable properties, such as similar sizes, bulk densities, and surface ice composition (3-5). As a result, they also share the same class of planetary atmosphere: at their distance from the Sun, their surface ices (nitrogen $N_2$, methane $CH_4$ and carbon monoxide CO) are indeed volatile enough to form a tenuous (~1 Pa) and condensable $N_2$ atmosphere with traces of $CH_4$ and CO. Through relaxation to solid-gas equilibrium, surface-atmosphere interactions take place, showing up in a variety of phenomena ranging from the formation of glaciers and icy dunes to the establishment of complex volatile cycles and climate systems. This distinct class of atmosphere shares similarities with Mars (tenuous and condensable) and Titan ($N_2$-$CH_4$ atmosphere with photolytically-produced hazes). It may be representative of other atmospheres which could form near perihelion on large volatile-rich KBOs (e.g., Eris, Makemake).

The appearances of Pluto and Triton are nevertheless very different. A first striking contrast between the two is the much smaller topographic extremes on Triton, less than ±1 km compared with ±4 km on Pluto (7,8). Second, Triton's surface is uniformly youthful, featuring very few impact craters (9,10), which is suggestive of global resurfacing. Pluto, too, has some youthful terrains that are evidently being modified by ongoing activity, but it also has widespread, ancient, heavily-cratered terrains that are inconsistent with any recent global-scale resurfacing process (11-13). Third, Pluto and Triton differ from the location of their main reservoir of volatile ice. On Pluto, the main reservoir of $N_2$ ice is confined within the massive topographic basin of Sputnik Planitia in the equatorial regions (11), while on Triton, $N_2$ ice seems to form an extended deposit in the southern hemisphere, possibly from 90°S to ~15°S (14). Fourth, Pluto exhibits dark, reddish equatorial regions, exemplified by Belton Regio (formerly Cthulhu Macula). The coloration of these regions is thought to result from complex organic haze particles that accumulate after settling to the surface (15). Once established, such regions appear to be able to resist accumulation of volatile ices, at least at low latitudes owing to their low albedos and attendant daytime temperatures (16,17). By contrast, Triton is uniformly bright (18) and does not display any dark tholin-covered terrains. Its $H_2O$-$CO_2$ bedrock is thought to be exposed over a large part of its surface (19,20). Finally, massive methane-rich deposits were observed on Pluto, in the form of mantled uplands (21) and "Bladed Terrain", a landform interpreted as analogous to terrestrial penitentes, albeit at a much larger spatial scale (22,23). By contrast, extended $CH_4$-rich deposits seem absent on Triton. This leads to abundances of gaseous $CH_4$ in Triton's atmosphere ~10 times lower than in Pluto's atmosphere (24,25). This also impacts the photochemical haze production (organic compounds originate from $CH_4$ UV photolysis) and the thermal structure (methane heats Pluto's atmosphere; haze particles also have a radiative impact (26) ). Ultimately, this likely explains the differences in the thermal structure of Pluto's and Triton's atmospheres, with (i) the lack of a marked stratosphere on Triton (27) and (ii) a warmer upper atmosphere (~90 K) on Triton vs Pluto (~70 K).

What could explain these differences? On the one hand, the geological history (and by extension, interior thermal history) of Pluto and Triton differ (2,28). It has been suggested that Triton has undergone and may still undergo subsequent resurfacing processes (e.g., cryovolcanism) due to intense internal deformation and heating powered by tidal activity with Neptune (6). This likely explains the uniform flatness and youth of its surface, and possibly associated features, not detected on Pluto, such as the cantaloupe terrain (29,30), the volcanic plains and calderas (30), the ridges and guttae (14) and the plumes erupting ~8 km up into Triton's atmosphere (31,32). This geological history could also have significantly impacted Triton's volatile cycles.

On the other hand, the variation of astronomical parameters (obliquity, eccentricity, solar longitude of perihelion) over time are known to be key drivers of solar insolation and surface temperatures on Earth, Mars and Titan. These parameters control the length and magnitude of the seasons and force the volatiles that form glaciers, lakes and frost to migrate through time. Previous modeling studies showed that Pluto's and Triton's climate are strongly sensitive to these variations (33-35), which drive significant changes in sublimation and condensation rates of the volatile ices and affect the atmospheric pressure and composition, and the surface distribution of volatile ices (35-38). However, these models did not explicitly address the differences between Pluto and Triton.

## The Pluto-Triton volatile transport model

Here we use the Pluto-Triton volatile transport model (VTM) of the Laboratoire of Meteorologie Dynamique (LMD) to compare in detail the impact of the obliquity and orbital parameters on volatile transport on Pluto and Triton (35,38). We perform long-term simulations of Pluto and Triton volatile cycles assuming the same model and initial conditions. Alternative simulations highlighting the impact of the obliquity and orbital parameters are also presented. The VTM has been developed on the basis that Pluto's and Triton's atmospheres have a negligible radiative thermal influence on the energy balance of their surface (see Materials and Methods). The model includes the calculation of the local insolation at the surface and the thermal infrared cooling, the heat storage and conduction in the subsurface, and, in the presence of nitrogen ice, glacial flow and the condensation-sublimation rates necessary to force the surface temperature to remain at the ice frost point. Methane is sublimed and condensed depending on the overlying vapor pressure and redistributed by a parameterized atmosphere (see further details in Materials and Methods). With these processes, the model is able to estimate the evolution of surface temperatures, pressure and volatile transport over seasonal and astronomical timescales on Pluto and Triton (35,38).

On Pluto, the orbital and obliquity parameters evolve on a timescale of thousands of Earth years, similar to the Milankovitch cycles on Earth. Pluto's obliquity (current: 119.6°) varies between 104° and 127° every 2.8 million years (Myr), and is therefore relatively high (its mean obliquity is 115°, i.e., 65° in retrograde rotation). Pluto's solar longitude of perihelion (current: 3.7°) varies from 0° to 360° with a precession period of 3.7 Myr, and its eccentricity (current: 0.2488) oscillates between 0.222 and 0.266 with a 3.95-Myr period (39-41). With such high obliquities, Pluto experiences highly pronounced seasons with long polar nights lasting for several Earth-decades and reaching very low latitudes. Furthermore, the generally high eccentricity of Pluto results in a seasonal North-South asymmetry in incoming insolation. The solar longitude at perihelion (the Pluto-Sun angle at perihelion, measured from the northern Hemisphere spring equinox) controls the duration of the season and thus this North-South asymmetry by favoring a more intense and shorter northern summer compared to the southern summer when perihelion solar longitudes are close to 90° and the opposite when values are close to 270°.

The orbital and obliquity parameters and the seasons of Triton are more complex to define because Triton orbits Neptune. Changes in inclination, precession period and Neptune obliquity are significant over a timescale of tens of millions of years and the Triton-Sun distance has not evolved much since its capture by Neptune (42), therefore the seasons on Triton over the last millions of years can be reconstructed from its current orbit around Neptune only. Triton has a retrograde and circular orbit around Neptune with an orbital inclination of 157°. The combination of this inclination with Neptune's obliquity of 28° leads to irregular seasons, with summer solstices oscillating between low (~5° latitude), moderate (~20° latitude) and extreme (~50° latitude) over a period of 650 Earth years, each season lasting ~35–45 Earth years and an annual-like cycle lasting ~140–180 Earth years (33,35). These oscillations are equivalent to a "planet-like obliquity" (i.e., the angle between Triton's rotation axis and Neptune's orbit plane) varying between 5°, 20° and 50° over a period of 650 Earth years, which corresponds to an average of ~30° over several Myr. Triton is therefore an object with a relatively low mean obliquity.

## Model results

Figure 1 shows modeled zonal mean surface temperatures on Pluto and Triton averaged over 10 Myr, in an ideal case of a volatile-free water ice bedrock. On average over such a timescale, surface temperatures are predicted to be slightly warmer at the poles than at the equator on Pluto by ~1 K, and colder at the poles than at the equator on Triton by 6-7 K (Figure 1.A). In these conditions, it is logical to assume that volatile ices would accumulate at the poles on Triton, and in the equatorial regions on Pluto, although one should note that the pole-to-equator temperature difference is small on Pluto. Figure 1.A also shows the modeled mean surface temperatures obtained on Pluto assuming a Triton-like obliquity. In this "low obliquity Pluto" case, surface temperatures become colder at the poles that at the equator, by 5 K. Figure 1.B shows the case of a circular orbit on Pluto as well as orbits with constant solar longitude of perihelion, thus highlighting the impact of these parameters on Pluto's surface temperatures. Variations between the different cases are in the order of 1-2 K only. In all cases, the poles remain warmer on average than the equatorial regions. The solar longitude of perihelion creates a North-South asymmetry in surface temperatures, favoring a warmer southern hemisphere when values are close to 90° and a warmer northern hemisphere when values are close to 270° (Figure 1.B). The surface temperatures are the same in both hemispheres for values close to 0° and 180 °.

These modeling results of the surface temperature hint at obliquity being the main driver for the ice distribution observed on Pluto and Triton. In order to investigate this in more detail, we performed simulations of Pluto and Triton with the LMD Volatile Transport Model spanning the last 20 Myr. We used a flat topography for Triton, and the New Horizons topography data for Pluto. In addition, we designed two alternative Pluto simulations with the following modifications: (a) the first alternative simulation assumes that Pluto orbits the Sun as the Neptune-Triton system does, i.e., at 30.1 au with an eccentricity of 0.009 (the obliquity remains that of Pluto), and (b) the second one assumes that the subsolar point follows the same variations with latitude and time as on Triton. We also explored alternative Triton simulations in which topography is not flat but instead includes a Sputnik-like 3-km deep region-scale topographic basin. All simulations start with a global and uniform cover of 200 m of $N_2$ ice and 4 m of $CH_4$ ice and share the same initial state for the surface and subsurface temperatures (uniform at 37 K). Only the topography, orbit and obliquity differ in these simulations.

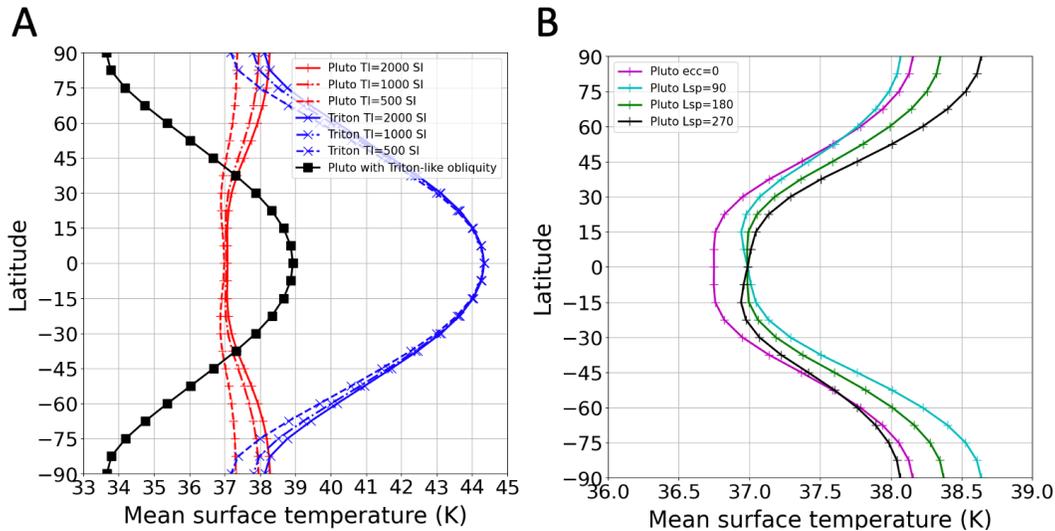

**Figure 1.** Modeled zonal mean surface temperatures on Pluto and Triton averaged over 10 Myr. A. For a surface with uniform albedo of 0.6, emissivity of 0.8, and subsurface thermal inertias of 2000, 1000, and 500 SI (see Materials and Methods). The black curve shows the mean surface temperatures on a virtual Pluto assumed to be on a Pluto-like orbit but with a Triton-like (i.e., low) obliquity. B. Same but assuming Pluto with a circular orbit (eccentricity=0, magenta) and constant solar longitude of perihelion of 90° (cyan), 180° (green), 270° (black).

## The role of obliquity

Figure 2 shows the distribution of volatile ice obtained after 20 Myr, with $N_2$ and $CH_4$ ice deposits in an equilibrated state. The Triton simulation shows that the volatile ice accumulates at the poles, forming massive polar caps of $N_2$-rich ice (Figure 2.B). The Pluto simulation reproduces the formation of perennial deposits of $N_2$ and $CH_4$ volatile ices in the equatorial regions, with $N_2$ ice in the Sputnik Planitia topographic basin and in local depressions outside, in line with New Horizons observations. This distribution is not strongly impacted by the change in eccentricity: if Pluto was orbiting the Sun in a quasi-circular orbit at the distance of Neptune (alternative simulation 1), the volatile ice would still accumulate in the Sputnik Planitia basin, as shown by Figure 2.C. On the other hand, if we consider a Pluto-like orbit with a Triton-like obliquity (alternative simulation 2), then the volatile ice would accumulate at the poles, as shown by Figure 2.D, and in line with the Voyager 2 observations. These results demonstrate that obliquity is the main driver explaining the differences in perennial volatile ice distribution observed on Pluto and Triton.

## The role of topography

Topography also plays an important role to trap the nitrogen ice at the longitudes of the deepest depressions (17). This is because the surface pressure – mostly $N_2$ – is higher at low elevation and so therefore is the equilibrium $N_2$ ice temperature, as dictated by the solid-gas equilibrium. It follows that the warmer $N_2$ ice deposits at low elevation emit more thermal infrared heat to space than the colder deposits at high altitude, which leads to increased $N_2$ condensation rates at low elevation allowing this difference in thermal radiation to be balanced by latent heat exchange. Consequently, $N_2$ ice tends to accumulate at low altitude, which explains why the Sputnik Planitia basin is filled with N$_2$ ice (17,43). This effect is particularly strong on Pluto, because differences between polar and equatorial annual mean temperatures are small, and therefore N$_2$ ice deposits at low elevation can be more stable than N$_2$ ice located in the coldest regions. On Triton, even with a deep basin this effect is limited because the poles are much colder than the equatorial regions on average. In our Triton simulations, if we place a Sputnik-like topographic basin in the equatorial regions, N$_2$ ice does not accumulate at all in it because the warm equatorial temperatures of the bedrock overcome the low-altitude effect (see Supplementary Figure 2). However, if the basin is placed at higher latitudes, Triton's ice distribution is slightly impacted as N$_2$ ice flows from higher to lower latitudes and at the bottom of the basin, where it becomes stable over long-term periods (Figure 2.E). The same effect is obtained in the simulation considering the Pluto-like orbit with a Triton-like obliquity (Figure 2.D), in which N$_2$ ice deposits become stable in the northern part of Sputnik Planitia, aside from the poles.

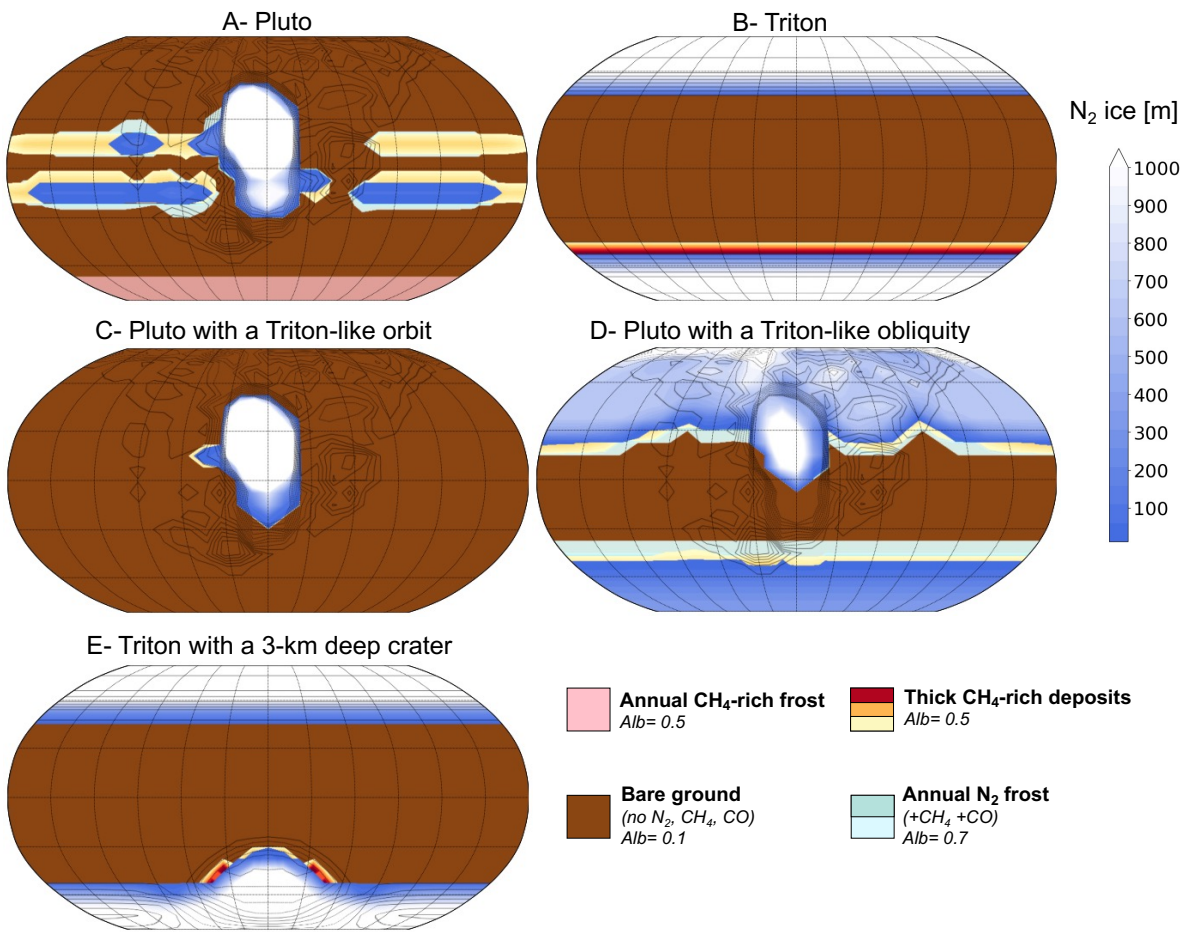

**Figure 2.** Volatile ice distribution obtained after a 20-Myr simulation performed with the LMD volatile transport model for (A) Pluto, (B) Triton, (C) Pluto with a Triton-like orbit around the Sun (eccentricity = 0.009, Sun-Triton distance = 30.1 AU) and with a Pluto-like obliquity (varying over a timescale of 2.8 Myr, see Section 2.1), (D) Pluto with a Pluto-like orbit but Triton-like obliquity (i.e, Triton's subsolar latitude variation), and (E) Triton with a 3-km deep (Sputnik-like) topographic basin centered at 60°S. Black contours show the bedrock's topography on Pluto (Triton is assumed flat).

Topography also impacts the $CH_4$ ice distribution. In the equatorial regions of Pluto, $CH_4$-rich ice has mostly been detected at high altitudes, for instance in the high-elevated Bladed Terrains and at the summits of mountain tops (44). A first explanation for this phenomenon is related to the altitude segregation with $N_2$-rich ice: $N_2$ tends to sublime at higher altitude and to leave $CH_4$-rich ice behind, which then becomes dominant at high elevations (23,45,46). A second explanation supported by climate models of Pluto suggests that the atmosphere is enriched in gaseous $CH_4$ at ~4 km altitude and that the highest mountains extend into this $CH_4$-enriched air and therefore get capped by $CH_4$ frost (47). This process of high-altitude $CH_4$ ice accumulation is further amplified by positive albedo feedback. These mechanisms are believed to work on Triton as well, although to a lesser degree since the topography is much less pronounced than on Pluto. Besides, the flatness of Triton may prevent the segregation by altitude of $N_2$-rich and $CH_4$-rich ice and contribute to the limited presence of $CH_4$-rich ice on Triton (48). In order to illustrate the topography effect on the $CH_4$ ice distribution on Pluto, we perform an alternative simulation of Pluto in which the gaseous $CH_4$ abundance is increased at high altitude (see Materials and Methods). The simulation now reproduces the formation of perennial deposits of $CH_4$ ice at high altitude in the equatorial regions, as observed by New Horizons (Figure 3.A).

The obliquity and topography on Pluto and Triton not only impacts the location of their surface volatile ice distribution but also affects their climate at a global scale. This is because (a) the distribution and seasonal evolution of surface $N_2$ ice reservoirs drive the $N_2$ condensation-sublimation flows which strongly controls the winds and the general atmospheric circulation (49-51), and (b) the location of the coldest regions may favor (in the case of Pluto) or prevent (in the case of Triton) the formation of extended $CH_4$-rich ice deposits (see below). On Triton, this absence of massive, perennial and extended $CH_4$-rich deposits leads to lesser amounts of gaseous $CH_4$ in the atmosphere than on Pluto, which causes colder atmospheric stratospheric temperatures and less evolved photochemical processes and haze formation. As a result, there are also less haze particles settling on the surface on Triton. This contributes to maintain the young surface bright along with the resurfacing processes, while on Pluto, the long-term accumulation of haze particles settling on the surface formed the dark equatorial regions.

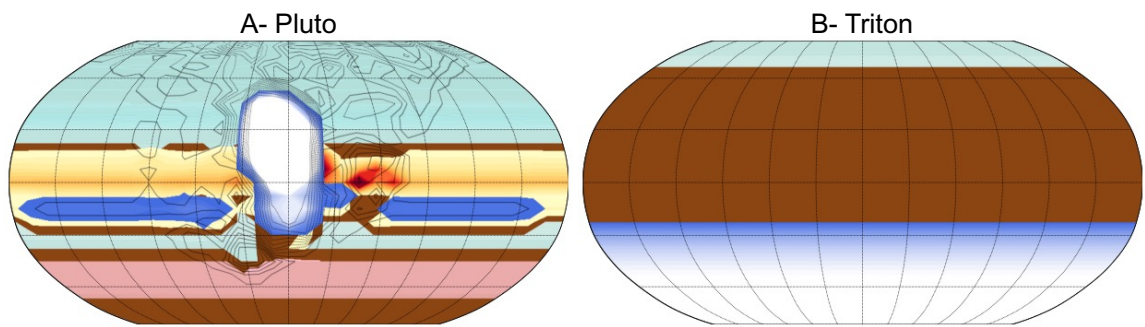

**Figure 3.** As Figure 2.A-B but for an alternative simulation of (A) Pluto with increased gaseous $CH_4$ abundance in the equatorial regions, and (B) Triton with a North-South $N_2$ ice albedo asymmetry.

How does the location of the coldest region impact the amount of $CH_4$-rich ice exposed at the surface? Pluto's low-insolation equatorial regions correspond to an extended area with variegated topography (± 5 km) allowing the existence of both perennial $N_2$-rich and $CH_4$-rich deposits, as $N_2$ preferably accumulates in depressions. On Triton, however, the low-insolation areas are the poles, which correspond to smaller areas displaying only small variations in topography (at least on the pole imaged by Voyager 2 (8)). As a result, on Triton, when $N_2$ condenses at the winter pole, it likely does so covering the entire polar region, leaving no space for exposed $CH_4$-rich ice deposits over large areas (35). It could still be possible that a lag deposit of $CH_4$-rich ice forms at the summer pole as $N_2$ sublimes (48), or that previously buried $CH_4$-rich ice deposits become exposed if the entire polar $N_2$ ice reservoir sublimed. However $N_2$ condensation in the winter season would then likely bury $CH_4$-rich ice again, similarly to the $H_2O$ and $CO_2$ ice caps on Mars. Another explanation for the lack of extended $CH_4$-rich deposits at Triton could also be a depletion in the total $CH_4$ abundance driven by internal chemistry or capture history. Current volatile transport models of Triton are not sensitive to the total abundance of $CH_4$ ice because, in the simulations, any initial $CH_4$ ice reservoir would migrate to the cold poles and always remain there, buried by $N_2$ ice deposits and therefore not interacting with the atmosphere. However, these models only treat $N_2$ and $CH_4$ as almost pure ice, and do not treat the case of intermediate mixtures, which may play an important role.

## Other processes impact Pluto and Triton landscapes

Obliquity and topography drive first-order differences in surface appearance on Pluto and Triton. Nevertheless, complex surface-atmosphere interactions leading to feedback between the ice properties and the condensation-sublimation rates over different timescales could also play an important part in disrupting the landscapes, and drive second-order differences on Pluto and Triton. For instance, in the equatorial regions of Pluto, runaway forcing by albedo are efficient mechanisms to maintain albedo contrasts over time (16,51) and therefore the presence of dark volatile-free features, methane-rich ice deposits and nitrogen-rich glaciers in proximity. The feedback processes could involve changes in grain size or composition due to sublimation and condensation (52), aeolian activity (51,53) and surface ice darkening through contamination by settling haze particles (51) or through direct irradiation by solar UV and cosmic rays (54,15), ultimately impacting the surface albedo. On Triton, North-South asymmetries in volatile ice cap extent have also been suggested to be caused by runaway albedo feedback (35,55). The same feedback process as on Pluto may be at play, as well as surface ice darkening due to geyser activity. Changes in $N_2$ ice phase and emissivity (56-58), or anomalies in internal heat flows locally impacting the thermal surface balance have also been suggested (59). Figure 3.B shows an alternative simulation of Triton in which the southern $N_2$ ice is brighter than the northern $N_2$ ice (see Materials and Methods). Because of this North-South albedo asymmetry, the southern cap becomes the main volatile reservoir (only seasonal frosts form in the northern hemisphere). The southern cap extends further toward the equator than for the reference Triton simulation, which is in better agreement with Voyager 2 observations.

## Conclusions

In conclusion, there are four main reasons that combine together to explain why landscapes of Pluto and Triton are markedly different. First, as shown in previous studies, Triton is tidally activated by Neptune while Pluto is not, and this led to intense internal activity on Triton and subsequent resurfacing processes, thus explaining Triton's flatness. Second, as shown here, Triton's mean obliquity with respect to the Sun is relatively low. As a result the poles are on average much colder than the equatorial regions and therefore accumulate volatile ices. On the contrary, the high obliquity on Pluto leads to volatile ice accumulation in the equatorial regions. The impact of the obliquity goes beyond the ice distribution and also explains key atmospheric and climate differences between Pluto and Triton. This is because the location of the coldest regions may favor (in the case of Pluto) or prevent (in the case of Triton) the formation of extended $CH_4$-rich ice deposits, which in turn impacts the atmospheric methane abundance, and subsequently, the atmospheric temperatures, photochemical processes and haze formation, darkening of the surface by settling haze particles, escape rates and climate at a global scale. Obliquity is therefore the key driver of Pluto's and Triton's different appearances. Third, topography triggers accumulation of $N_2$-rich ice at low elevation (i.e., in Sputnik Planitia) and $CH_4$-rich ice at altitude (e.g., the Bladed Terrains), and this is particularly pronounced in the equatorial regions of Pluto because of the dramatic topography combined to the high obliquity. Finally, complex surface-atmosphere interactions can locally impact the surface balance and lead to runaway albedo feedback. On both Pluto and Triton, these feedbacks are efficient in the equatorial regions, thus driving longitudinal contrasts in surface ice distribution and further disrupting the landscapes.

The fact that we can reproduce the observed surface ice distributions that are at the origin of the different landscapes and climates of Pluto and Triton, by using the same model and initial conditions, gives support to the hypothesis that both objects had a common origin, and later a different dynamical history resulting from differences in obliquity and topography.

## Materials and Methods

### The LMD Pluto-Triton Volatile Transport Model (VTM)

We used the latest version of the Pluto-Triton volatile transport model (VTM) of the Laboratoire de Météorologie Dynamique (LMD) (35,38). The VTM is a 2D surface thermal model, which computes at each point the surface radiative budget, the exchange of heat with the subsurface by conduction, and the phase changes and the exchanges of volatile ($N_2$, CO, $CH_4$) with the atmosphere, and takes into account a glacial viscous flow scheme for $N_2$ ice (37,60), and the seasonal and astronomical variation of the subsolar point specific to Pluto and Triton (see below). Here the CO cycle is neglected, because in the model CO always follow $N_2$ and CO ice is always mixed into the $N_2$-rich ice. We consider that both Pluto and Triton's atmosphere are very tenuous and almost transparent and thus have a negligible influence on the surface thermal balance aside from the condensation-sublimation exchanges of latent heat with the surface. We parametrize the atmospheric transport using a simple global mixing function for $N_2$ and $CH_4$ in place of 3D atmospheric transport and dynamics, with a characteristic time τ for the redistribution of the surface pressure and trace species, based on reference 3D global climate model simulations (51). Tests done with the 3D global climate model determined the timescales for atmospheric transport of $CH_4$ ($10^7$ s, i.e., about 4 Earth months) and $N_2$ (1 s, instantaneous mixing) used in the VTM (values are the same for both Pluto and Triton (35,38)). The topography in the model is controlled by the amount of volatile ice on the surface. For Triton we use a flat topography for the non-volatile bedrock; for Pluto we use the latest topography from New Horizons data (7) with flat topography for the non-observed southern hemisphere. Adding topography in Pluto's southern hemisphere does not impact the results of this paper.

Note that the numerical model only applies to global atmospheres. For Pluto, the approximate limit at which point the atmosphere is non-global is 0.006 Pa (61,62). For Triton, the rescaled threshold is 0.009 Pa (owing to the larger size and gravity constant). In this paper, the pressure generally remains above 10 mPa for both Pluto and Triton in all simulations. A few simulations reach the threshold but over an extremely short range of time, and therefore we do not expect this process to impact the results significantly (in addition these low pressures involve very slow sublimation and condensation rates and the ice distribution is therefore not impacted). Overall, this implies that Pluto's and Triton's atmospheres remain global over short-term and long-term periods, as shown in previous studies(35,37).

Simulation settings

Pluto and Triton surfaces are represented by a grid of 32 longitudes x 24 latitudes. All simulations start with a global and uniform cover of 200 m of $N_2$ ice, and 1 m of $CH_4$ ice. They are run over the last 20 millions of Earth years, which is long enough for the surface and subsurface to reach a steady state insensitive to the initial state. To do that we use a paleoclimate and ice equilibration algorithm (37). The model is first run over several Pluto or Triton years to capture several annual cycles so that the ice distribution and surface pressure reach an equilibrium. We consider that the first annual cycles correspond to a spin up time and we use the last seasonal cycle to estimate the mean sublimation-condensation rates over that period. These annual mean sublimation-condensation rates are then extrapolated over a paleo-timestep of Δt=20,000 Earth years to calculate the new amounts of surface volatile ice and the corresponding changes in topography.

Note that for the simulations of Pluto, the paleo-timestep must be short enough so that the changes in obliquity and orbital parameters allow the surface ice distribution to reach a steady state at each paleo timestep (but it must be long enough to reduce significantly the computing time of the simulation, as a trade-off). Also, any change in obliquity over a paleo-timestep must be smaller than the latitudinal resolution of the model. In previous work (Bertrand et al., 2018), we used a paleo-timestep of ~50,000 Earth years, which corresponds to about 200 Pluto orbits and a maximal change in its obliquity of 1° of latitude (Binzel et al., 2017). For the simulations of Triton, the paleo-timestep is driven by the timescale of the glacial flow since the obliquity varies over a timescale of hundreds of Earth years and associated surface changes are covered by the simulation. In previous work (Bertrand et al., 2022), we used a paleo-timestep of ~20,000 Earth years. In this paper, for sake of consistency, we used a paleo-timestep of ~20,000 Earth years for both Pluto and Triton. We ensured that the steady state is reached at each paleo-timestep.

The simulations are performed using an $N_2$ ice bolometric emissivity of 0.8 and bolometric albedo of 0.7. The surface $N_2$ pressure simulated in the model is constrained by these values. Note that our model reproduces very well the pressure evolution observed on Pluto (17,63) and relatively well that observed on Triton (35).The albedo and emissivity of the bare ground (volatile-free surface) are set to 0.1 and 1 respectively. Methane ice bolometric albedo and emissivity are set to 0.6 and 0.8, respectively, in all simulations, based on observations (64,65). The thermal conduction into the subsurface is performed with a low thermal inertia near the surface, set to $I_d$=20 J s$^{-0.5}$ m$^{-2}$ K$^{-1}$, to capture the short-period diurnal thermal waves and a larger thermal inertia below set to $I_s$=800 J s$^{-0.5}$ m$^{-2}$ K$^{-1}$, to capture the much longer seasonal thermal waves which can penetrate deep in the high-TI substrate. The rest of the settings are similar to the previous simulations of Pluto and Triton with the LMD VTM (35,38).

### Pluto and Triton obliquity and orbital parameters

The Pluto simulations calculate the obliquity and orbital (solar longitude of perihelion and eccentricity) changes over time (40). The incoming insolation therefore varies with the 2.8 Myr obliquity cycle and the 3.7 Myr cycle of eccentricity and libration of argument of perihelion. The Triton simulations compute the subsolar latitude on Triton from the dynamical solution developed by Forget et al.(35,66). These equations remain valid over the last 20 Myrs, assuming no change over time in obliquity (with respect to Triton's orbit around Neptune) and orbital parameters, which is reasonable over this timescales. The final years of the simulation correspond to Pluto's and Triton's current epoch.

### $CH_4$ condensation-sublimation on the surface

$CH_4$ is a minor constituent of Pluto's and Triton's $N_2$ atmosphere and the surface-atmosphere interactions depend on the turbulent fluxes given by:

$$F = \rho\, C_d\, U\, (q - q_{surf}) \quad (1)$$

Where $q_{surf}$ is the saturation vapour pressure mass mixing ratio (in kg kg$^{-1}$) at the considered surface temperature, $q$ the atmospheric mass mixing ratio, $\rho$ the air density, $U$ the horizontal wind velocity (set to 0.5 m s$^{-1}$ as suggested by both the Pluto and Triton 3D models) and $C_d$ the drag coefficient at 5 m above the local surface ($C_d$=0.06). $q_{surf}$ is computed using the thermodynamic Claudius-Clapeyron relation (67) for $CH_4$ with a latent heat for sublimation of 586.7 kJ kg$^{-1}$.

Under specific thermodynamic assumptions, volatile ices on Pluto's and Triton's surface should form solid solutions whose phases follow a ternary phase diagram (45,68). The ices should relax to a 3-phase equilibrium in which the sublimation pressure of each species depends only on temperature and is independent of the $CH_4$ and CO mole fraction. However, the prediction of the 3-phase system at equilibrium does not correspond to what has been seen by New Horizons, which detected a large diversity of ice mixtures out of equilibrium. Pluto and Triton are rather interpreted to be non-equilibrium dynamical environments with continuous exchange of materials. In this context and awaiting more sophisticated schemes for simulating the behaviour of the different ice mixtures, we use Raoult's law as a simplification of the ternary phase diagram. The CO cycle is neglected (CO tends to follow $N_2$), and we consider the mixture $N_2$:$CH_4$ with 0.5% and 0.05% of $CH_4$ on Pluto and Triton respectively, as retrieved from telescopic observations and from New Horizons observations (48,69).

### Summary of the simulations presented in this paper

The reference simulations of Pluto and Triton are shown in Figure 2.A and Figure 2B, respectively. Alternative simulations for Pluto include (1) a simulation with a quasi-circular orbit (assuming an eccentricity of 0.009 as for Neptune) at the distance of Neptune, Pluto with a Triton-like orbit, Figure 2.C), (2) a simulation with a long-term evolution of the sub-solar point as on Triton (Pluto with a Triton-like obliquity, Figure 2.D), and (3) a simulation in which the near-surface gaseous $CH_4$ abundance in the equatorial regions depends on the altitude, as suggested by 3D climate model results (51) (Figure 3.A). The gaseous $CH_4$ abundance is artificially increased by a factor of 2 at ~4 km altitude compared to the gaseous $CH_4$ abundance at the mean surface level, and follows a linear function between these two levels.

Alternative simulations for Triton include (a) a simulation with a 3-km deep Sputnik-like basin at 60°S (Figure 2.E), (b) a simulation with brighter southern $N_2$ deposits than the northern deposits (the $N_2$ ice albedo is set to 0.75 and 0.65 in the southern and northern hemisphere, respectively, Figure 3.B), and (c) simulations performed with a 8-km deep Sputnik-like basin placed at different latitudes (from 0° to 60°S, Supplementary Figure 2).

## Acknowledgments

We thank N. Descamp, J. Berthier and C. LeGuyader for their previous work on calculating the ephemerides of solar system bodies and in particular those of Triton. We thank W. Grundy and L. Young for valuable discussions and comments on the topic of this manuscript.